\documentclass[aps, 12pt, letterpaper,showpacs]{revtex4-1}

\usepackage{amsbsy,amssymb,amsmath,bm, gensymb}
\usepackage{graphicx,color,epsfig,rotate}

\begin{document}

\title{The role of stoichiometric vacancy periodicity in pressure-induced amorphization of the Ga$_2$SeTe$_2$ semiconductor alloy}

\author{N.M. Abdul-Jabbar$^{1, 2}$, B. Kalkan$^3$, G-Y. Huang$^4$, A.A. MacDowell$^3$, R. Gronsky$^5$, E.D. Bourret-Courchesne$^2$, B.D. Wirth$^{1,4}$}
\affiliation{$^1$Department of Nuclear Engineering, University of California, Berkeley, California 94720, USA}
\affiliation{$^2$Materials Sciences Division, Lawrence Berkeley National Laboratory, Berkeley, California 94720, USA}
\affiliation{$^3$Advanced Light Source, Lawrence Berkeley National Laboratory, Berkeley, California 94720, USA}
\affiliation{$^4$Department of Nuclear Engineering, University of Tennessee, Knoxville, Tennessee 37996, USA}
\affiliation{$^5$Department of Materials Science and Engineering, University of California, Berkeley, California 94720, USA}

\begin{abstract}

We observe that pressure-induced amorphization of Ga$_2$SeTe$_2$ (a III-VI semiconductor) is directly influenced by the periodicity of its intrinsic defect structures. Specimens with periodic and semi-periodic two-dimensional vacancy structures become amorphous around 10-11 GPa in contrast to those with aperiodic structures, which amorphize around 7-8 GPa. The result is a notable instance of altering material phase-change properties via rearrangement of stoichiometric vacancies as opposed to adjusting their concentrations. Based on our experimental findings, we posit that periodic two-dimensional vacancy structures in Ga$_2$SeTe$_2$ provide an energetically preferred crystal lattice that is less prone to collapse under applied pressure. This  is corroborated through first-principles electronic structure calculations, which demonstrate that the energy stability of III-VI structures under hydrostatic pressure is highly dependent on the configuration of intrinsic vacancies.

\end{abstract}

\pacs{61.50.Ks, 61.05.cp, 62.50.-p, 81.30.Hd}

\maketitle

Ga$_2$SeTe$_2$ is a novel  III-VI semiconductor that exhibits a cubic zincblende crystal structure (F\={4}3\emph{m} space group) dominated by stoichiometric or ÒstructuralÓ vacancies (also known as defect zincblende). These defects arise due to the valence mismatch between the anion and cation forcing 1/3 of the cation sites to be vacant. Recent investigations on binary III-VI materials have shown that the presence of stoichiometric vacancies can lead to material properties crucial for phase change random access memory technology and thermoelectrics \cite{Sun:2006bo, Rasmussen:2013hz, Zhu:2010jj, Kurosaki:2008cj}. Similar to the binary III-VI compounds, Ga$_2$SeTe$_2$ may also have potential for phase-change memory applications. Its calculated average number of covalence electrons per single atom (N$_{sp}$) is 4.8, which meets a key criterion for successful phase-change materials \cite{Luo:2004vw, Lencer:2011bt}. Unlike the binaries, however, the ternary system may provide the capability of band-gap engineering, though one must be cognizant of solid-solution immiscibility in the Se rich region of the Ga$_2$Te$_3$-Ga$_2$Se$_3$ phase diagram \cite{Warren:1974ux}. Nevertheless, a thorough understanding of the phase-change dynamics of Ga$_2$SeTe$_2$ is a necessary first step in realizing its technological potential for memory applications.

Recent high-pressure x-ray diffraction experiments on Ge$_2$Sb$_2$Te$_5$ reveal that its cubic phase (which requires stoichiometric vacancies to be electrically stable) cannot be recovered after compression and decompression; conversely, its trigonal phase, where stoichiometric vacancies are absent, is preserved \cite{Krbal:2009in}. This is an impactful demonstration of the effect of the presence of vacancies on the pressure-induced properties of chalcogenide semiconductors. In this letter, we investigate the effect of applied pressure on Ga$_2$SeTe$_2$ and argue that in addition to their presence, the periodicity of stoichiometric vacancy structures in a crystal can influence pressure-induced material properties.

Ga$_2$SeTe$_2$ powders ($\approx$60 $\mu$m particle size) were prepared by crushing as-grown and annealed single crystals. Bulk single crystals were grown via a modified Bridgman technique using stoichiometric amounts of 8N Ga, 6N Se, and 6N Se \cite{AbdulJabbar:2012bma}. Synchrotron powder x-ray diffraction measurements at ambient conditions were performed at beam line 11-BM at the Advanced Photon Source (APS) at Argonne National Laboratory (ANL). Diffraction at high pressures and temperatures was performed at beam line 12.2.2 at the Advanced Light Source (ALS) at Lawrence Berkeley National Laboratory (LBNL). A focused beam (10 $\times$ 10 $\mu$m spot size) of 30 keV (0.4133 \r{A}) x rays was used to perform the diffraction experiments. Diffraction images were collected using a MAR 345 image plate detector. A LaB$_6$ standard powder specimen was used to measure sample-detector distance and detector tilt angles.

High temperature diffraction at ambient pressure involved mounting sealed (10$^{-6}$ Torr) 1 mm diameter quartz capillaries containing Ga$_2$SeTe$_2$ powder between two halogen lamps where the sample can be radiantly heated. Temperature was controlled by toggling the voltage supplied to the lamps and was measured using a type-K thermocouple. The voltage-current relationship was linear and temperatures up to 820 \degree C were accessible, sufficient for our investigations (Ga$_2$SeTe$_2$ melts at 800 \degree C).

High pressure x-ray diffraction measurements were performed at ambient temperature using a standard symmetric diamond anvil cell which consists of two 300 $\mu$m culet diamonds with a 60 $\mu$m indented rhenium gasket in between. Ga$_2$SeTe$_2$ powder mixed with pressure transmitting fluid (a 4:1 methanol/ethanol mixture) and a pressure marker (ruby pieces) was loaded in a 180 $\mu$m hole drilled at the center of the indentation. A more detailed description of the instrument setup at beam line 12.2.2 at the ALS has been reported elsewhere \cite{Kunz:2005ga}. High-pressure x-ray diffraction experiments conducted at high temperatures were done using a resistively heated diamond anvil cell. Liquid argon was used as the pressure transmitting fluid and gold was used as a pressure-temperature marker. Changes in volume of the gold pressure marker allowed for the measurement of the overall pressures in the experimental setup \cite{Heinz:1984vw}. Diffraction data were analyzed using the Celref software to determine unit cell parameters and volumes \cite{Cellref}.

Powder x-ray diffraction experiments conducted at ambient temperature and pressure reveal that the configuration of stoichiometric vacancies in Ga$_2$SeTe$_2$ is strongly related to specimen thermal history. Three different thermal treatments (analogous to those previously applied to Ga$_2$Te$_3$ \cite{Kim:2011kb}) were employed: (1) anneal sample at 735 \degree C for 28 days then quench to 0 \degree C, (2) anneal sample at 435 \degree C for 28 days followed by slow cooling in furnace, and (3) sample in its as-grown state. The diffraction pattern for 735 \degree C annealed specimen contains sharp ancillary peaks (labeled a and b) at $\Delta ($2$\theta$)=0.45\degree and $\Delta ($2$\theta$)=0.20\degree around the 111 Bragg reflection corresponding to a face-centered cubic Bravais lattice associated with a zincblende crystal structure with lattice constant $a$ = 5.77 \r{A} (Figure 1a). This alludes to additional superstructures parallel to $<$111$>$ directions. Their presence as two-dimensional vacancy structures has been verified via conventional and high-resolution electron microscopy \cite{AbdulJabbar:2014fl, Guymont:1992vj, Kurosaki:2008cj, Kim:2009dz}. In Ga$_2$SeTe$_2$, they order at 2.7 nm intervals or about 1/8 of the spacing of $\{111$\} planes in the zincblende lattice \cite{AbdulJabbar:2014fl}. Satellites become diffuse in the 435 \degree C annealed powder, signifying that the two-dimensional vacancy structures observed in the 735 \degree C annealed powder lose their periodicity. The diffraction pattern in the as-grown powder (previously reported elsewhere) is intermediate between the vacancy ordered and disordered samples \cite{AbdulJabbar:2012bma}.

To probe temperature dependence on vacancy periodicity in Ga$_2$SeTe$_2$, we collected x- ray diffraction patterns for the as-grown sample as a function of temperature at ambient pressure. Temperature was increased from ambient to 735 \degree C and then decreased back to ambient in 100 \degree C increments. At each temperature step, a diffraction pattern was collected after 300 s. Results are shown in Figure 1b. As the temperature is increased, prominent satellites begin to form around the 111 and 002 Bragg reflections. Upon cooling, the diffraction patterns show no change from the 735 \degree C state. This suggests that the mechanism driving the periodicity of two-dimensional vacancy structures in Ga$_2$SeTe$_2$ is not discrete, but a gradual process driven by temperature. Satellite reflections remain well defined upon fast cooling, which is expected based on the 735 \degree C annealing treatment applied to Ga$_2$SeTe$_2$ powders.

\begin{figure}
\centering
\includegraphics[width=12cm]{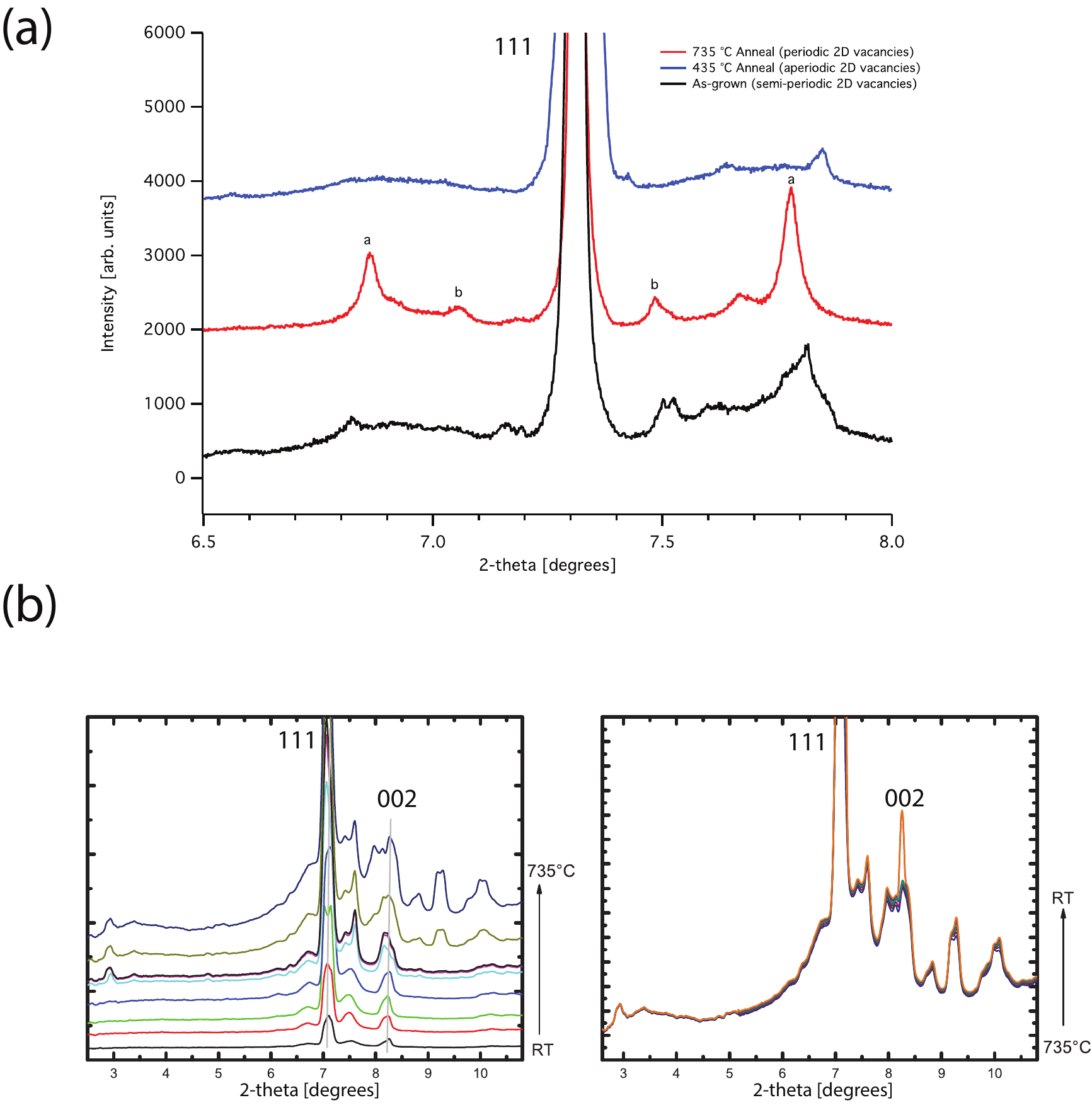}
\caption{(color online) (a) Diffraction patterns of as-grown (blacks), 735 \degree C (red) annealed, and 435 \degree C (blue) annealed Ga$_2$SeTe$_2$ powders at ambient conditions around the 111 Bragg reflection. Annealing at 735 \degree C and quenching to 0 \degree C produces sharp satellite peaks around the 111 Bragg reflection (labelled a and b) attributed to two-dimensional vacancy structures parallel to the $<$111$>$ directions. The satellites are still present in the as-gown and 435 \degree C specimens, but are diffuse, attributed to a decrease in the extent of vacancy periodicity. (b) Diffraction patterns of as-grown  Ga$_2$SeTe$_2$ as a function of increasing and decreasing temperature.}
\label{fig1}
\end{figure}

\begin{figure}[ht]
\centering
\includegraphics[width=12cm]{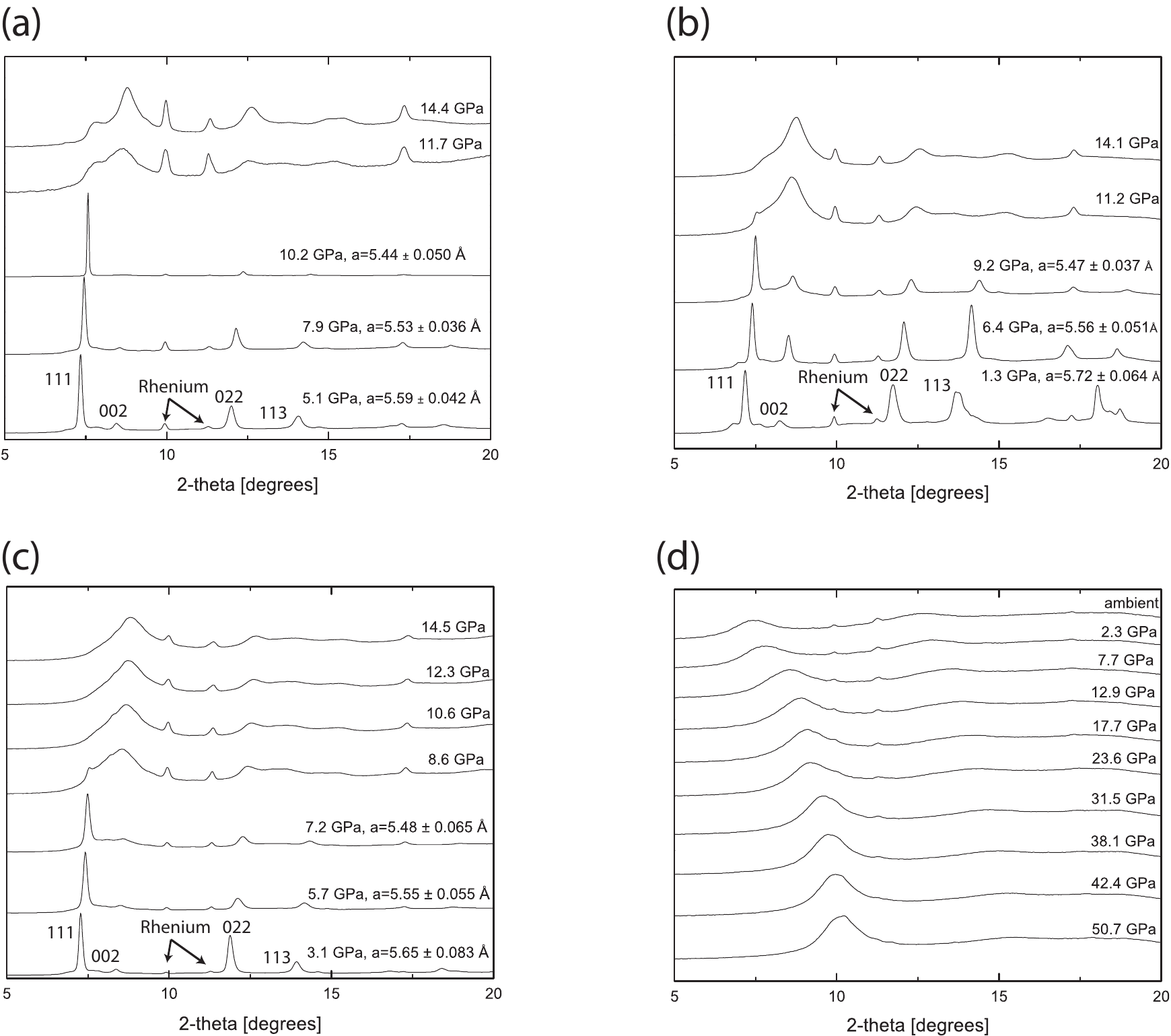}
\caption{X-ray diffraction patterns as a function of increasing pressure for (a) as-grown (semi-periodic two-dimensional vacancy structure), (b) 735 \degree C annealed (periodic two-dimensional vacancy structure), and (c) 435 \degree C annealed (aperiodic two-dimensional vacancy structure) Ga$_2$SeTe$_2$ samples. (d) X-ray diffraction patterns for the 735 \degree C annealed  Ga$_2$SeTe$_2$ upon decompression from 50 GPa to ambient conditions. After decompression the sample remains amorphous. Identical behavior was observed for the as-grown and 435 \degree C annealed specimens.}
\label{fig2}
\end{figure}

X-ray diffraction patterns as a function of increasing pressure at ambient temperature are shown for the as-grown, 435 \degree C, and 735 \degree C annealed Ga$_2$SeTe$_2$ specimens in Figures 2a-2c. All samples at ambient pressures exhibit a cubic zincblende structure. Additional peaks located at 2$\theta$=9.9\degree and 2$\theta$=11.3\degree are attributed to the rhenium gasket in the diamond anvil cell. As pressure is increased, the Bragg reflections shift toward higher angles, indicative of unit cell compression. During compression, the as-grown and 735 \degree C annealed specimens undergo a cubic to amorphous phase transition between 10 GPa and 11 GPa as revealed by the formation of broad amorphous bands at 2$\theta$$\approx$8.8\degree and 2$\theta$$\approx$12.5\degree. Above $\approx$12 GPa, the specimens remain amorphous with no additional phase transitions up to $\approx$50 GPa. The compression behavior of the 435 \degree C annealed sample is identical to the other two samples with the exception that solid-state amorphization occurred between 7 GPa and 8 GPa. Additionally, diffraction patterns upon decompression were collected for the as-grown, 435 \degree C, and 735 \degree C annealed specimens. For all samples, the cubic phase was not recovered (Figure 2d shows the decompression results for the 735 \degree C annealed sample). As vacancies electrically stabilize the cubic structure, such irreversibility is expected and has been previously observed in cubic Ge$_2$Sb$_2$Te$_5$ \cite{Krbal:2009in}.

Using the lattice parameters extracted from the high-pressure diffraction patterns and Birch-Murnaghan equation of state, bulk modulus values of 40.5 GPa, 45.0 GPa, and 38.0 GPa were computed for as-grown, 735 \degree C, and 435 \degree C annealed samples, respectively. This suggests that the periodicity of two-dimensional stoichiometric vacancy structures has a direct effect on the pressure-induced amorphization of Ga$_2$SeTe$_2$. To verify this observation, we performed diffraction experiments under high pressures and temperatures. We first heat the 735 \degree annealed sample (which exhibits periodic two-dimensional vacancy structures) to $\approx$100 \degree C and $\approx$170 \degree C followed by pressure-induced amorphization. As shown in Figure 3, the amorphization pressure decreases with increasing temperature. This implies a negative Clapeyron slope, or $d$$P$/$d$$T$=$\Delta$$S$/$\Delta$$V$, where $\Delta$$S$ and $\Delta$$V$ are the changes in volume and entropy respectively in going from the cubic to amorphous state. Since $\Delta$$S$$>$0 in this transition, the negative $P$-$T$ slope observed in the 735 \degree C annealed Ga$_2$SeTe$_2$ implies that its amorphous phase is denser than its cubic phase. Interestingly, as the 435 \degree C annealed sample (which exhibits aperiodic two-dimensional vacancy structures) is heated from ambient temperature, amorphization pressure increases and it adopts a pressure-temperature behavior similar to the 735 \degree C annealed specimen. Since vacancy periodicity in Ga$_2$SeTe$_2$ is induced by temperature (as seen in Figure 1b), the result affirms its role in increasing amorphization pressure (as seen Figure 2).

\begin{figure}
\centering
\includegraphics[width=\textwidth]{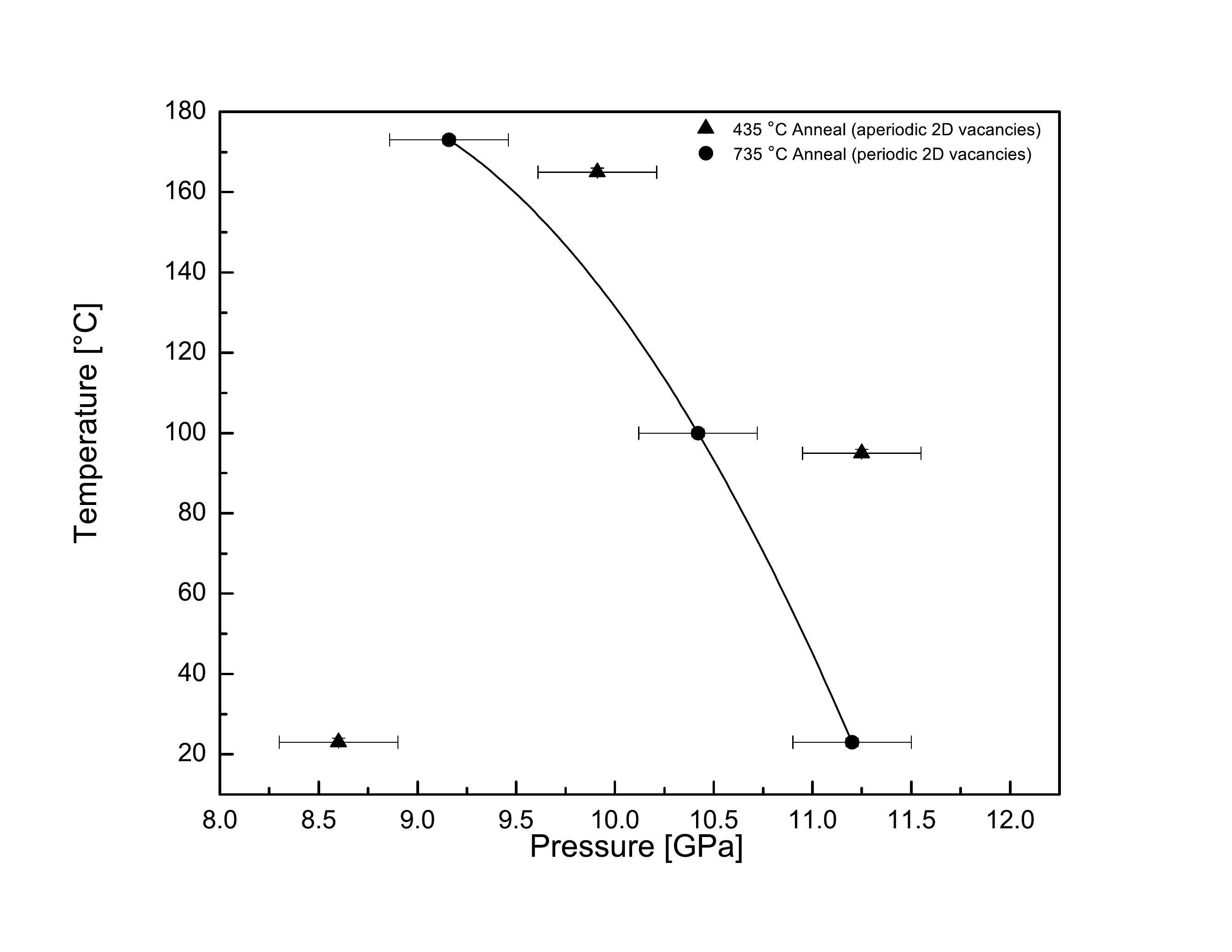}
\caption{Amorphization pressures for Ga$_2$SeTe$_2$ with periodic and aperiodic two-dimensional vacancy structures as a function of temperature.}
\label{fig3}
\end{figure}

\begin{figure}
\centering
\includegraphics[width=8cm]{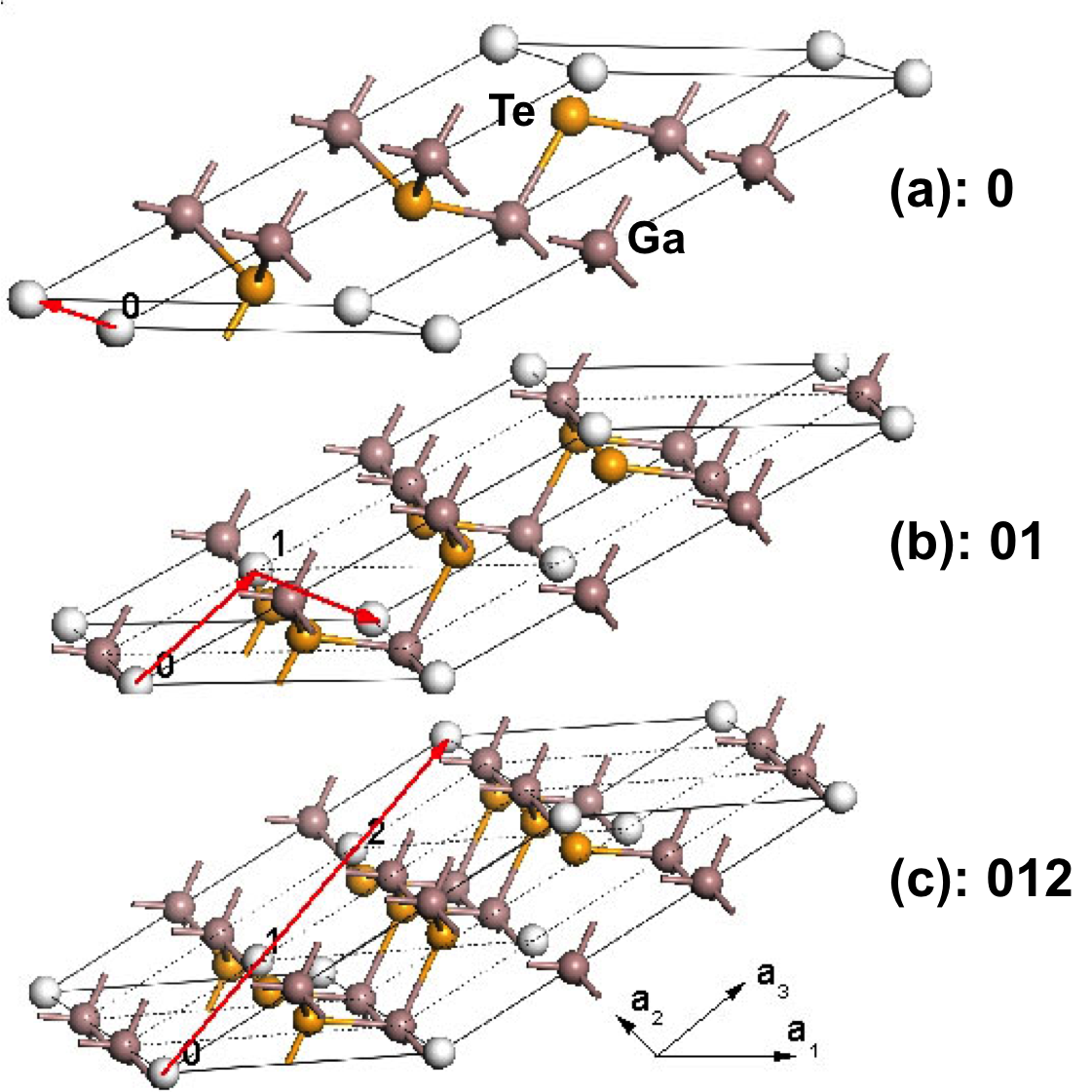}
\caption{(color online) Schematics of Ga$_2$Te$_3$-type structures with vacancies represented by white spheres and their configuration highlighted by a red line: (a) straight-line vacancy configuration with no lattice shift, (b) zigzag-line vacancy configuration with one lattice shift in the a$_3$ direction, (c) straight-line vacancy configuration with one lattice shift in first layer and two lattice shifts in the second layer along the a$_3$ direction.}
\label{fig4}
\end{figure}

\begin{figure}
\centering
\includegraphics[width=12cm]{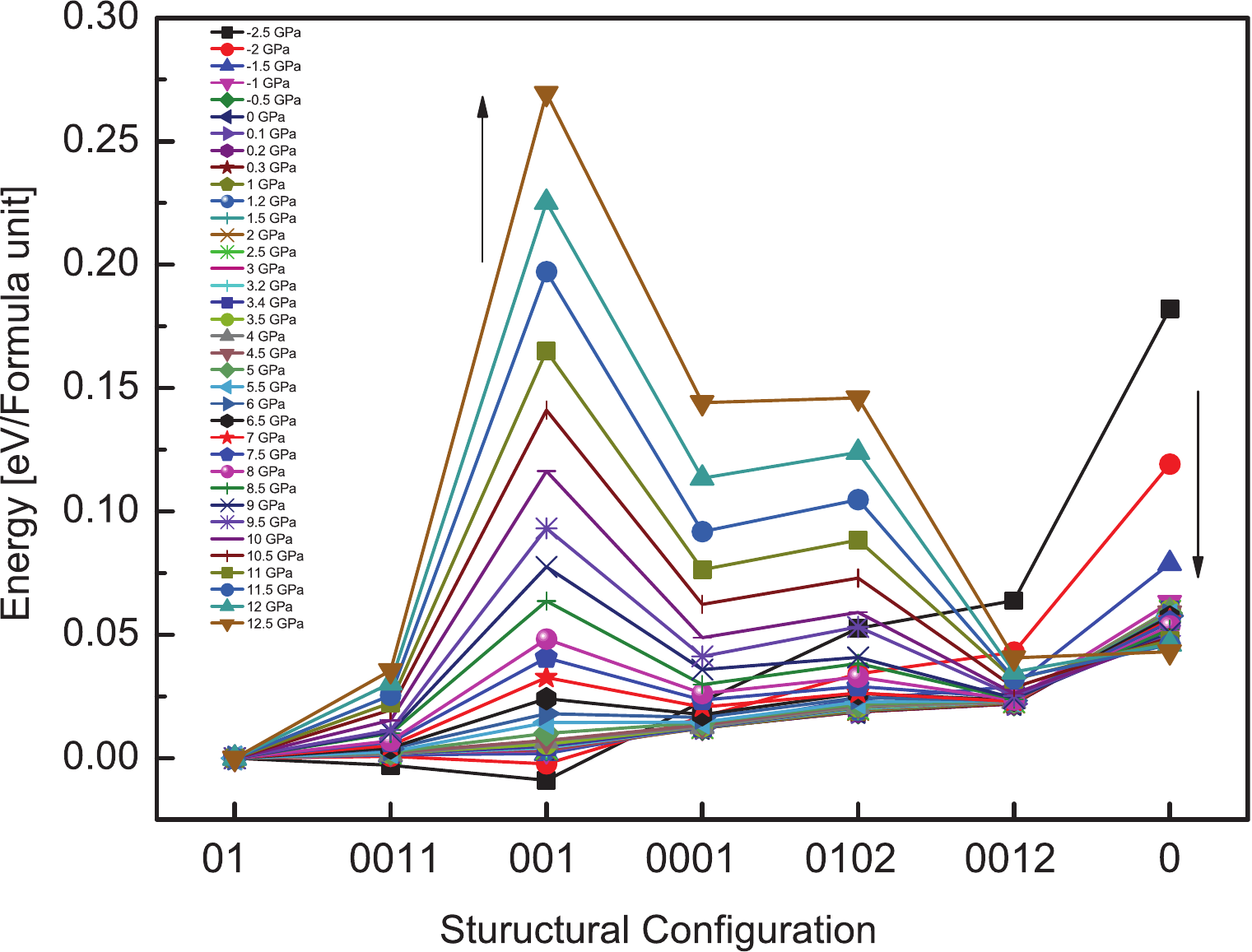}
\caption{(color online) Effect of hydrostatic pressure on the energies of various Ga$_2$Te$_3$ supercell structures.}
\label{fig5}
\end{figure}

The high pressure and temperature diffraction data collected for Ga$_2$SeTe$_2$ suggest that the arrangement of stoichiometric vacancy defect structures plays a role in the stability of crystal structure under an applied pressure. Recent theoretical work on the structure of Ga$_2$Se$_3$ has shown that its $\beta$-phase, comprised of a monoclinic crystal with an ordered zig-zag line vacancy superstructure, is the most energetically stable \cite{Huang:2013ke, Huang:2013jv, Huang:2014vt}. We perform density functional theory (DFT) calculations to examine the relationship between vacancy configuration and crystal stability under hydrostatic pressure using the Vienna \textit{ab initio} simulation package (VASP) utilizing the local density approximation (LDA) and the generalized-gradient approximation (GGAÐPBE) \cite{Perdew:1981dv, Kresse:1993bz, Kresse:1996kl}. We constructed a Ga$_2$Te$_3$ supercell with multiple vacancy configurations (denoted by a ternary numbering system shown in Figure 4a-4c). A binary III-VI system was studied in order to establish vacancy configuration as the sole variable. Figure 4a corresponds to a primitive cell with straight-line vacancy ordering (highlighted by the red line) that adopts an In$_2$Te$_3$ (orthorhombic) structure \cite{Woolley:1959tl}. This is denoted as 0 and means there is no shift along the a$_3$ direction. Figure 4b represents the primitive cell with zigzag-line vacancy ordering that adopts a $\beta$-Ga$_2$Se$_3$ (monoclinic) structure \cite{Lubbers:1982wd}. This is denoted by 01, which means there is no shift for the first layer and one lattice shift in the second layer along the a$_3$ direction (as denoted by the red line). In Figure 4c, a variation of the 0 structure with straight-line vacancies is shown. There is no shift in the first layer of the cell, one lattice shift along a$_3$ in second layer, and two lattice shifts along a$_3$ in the third layer; consequently this configuration is denoted as 012. One can adopt this structural convention to crystal systems of any size. 

Figure 5 shows the relative energies of Ga$_2$Te$_3$-type superstructures as a function of hydrostatic pressure. Four zigzag-line vacancy and three straight-line vacancy configurations are examined. The $\beta$-Ga$_2$Se$_3$ structure (01) has the lowest energy and remains constant with pressure. Zigzag-line vacancy configurations based on three and four cell superstructures (001, 0011, 0001) become less energetically stable with increasing pressure. Conversely, the energy of straight-line vacancy structures is lowered as pressure is increased (with the exception of of the 0102 configuration). The structural permutations that are available for analysis for III-VI materials are limitless, nevertheless, our theoretical calculations demonstrate that their stability under pressure can be highly dependent on vacancy arrangement---as we observed by our diffraction experiments.

We conclude that in the Ga$_2$SeTe$_2$ semiconductor alloy, there is considerable interplay between the periodicity of two-dimensional vacancy structures and pressure-induced amorphization. Observation of such a phenomenon provides further insight into the relationship between crystallographic defects and the physical properties of materials. Elucidating such a relationship is pivotal in gauging the technological potential of a large class of materials with structures dominated by stoichiometric vacancies.

The authors acknowledge C.A. Ramsey for experimental assistance and E.C. Samulon and G.A. Bizarri for useful discussions. N.M.A. acknowledges support from the Nuclear Nonproliferation International Safeguards Graduate Fellowship Program sponsored by the National Nuclear Security Administration's Next Generation Safeguards Initiative (NGSI). This work was supported by the US Department of Energy/NNSA/NA22 and carried out at the Lawrence Berkeley National Laboratory under Contract No.AC02Ñ05CH11231. Use of the Advanced Photon Source at Argonne National Laboratory was supported by the U. S. Department of Energy, Office of Science, Office of Basic Energy Sciences, under Contract No. DE-AC02-06CH11357. The Advanced Light Source is supported by the Director, Office of Science, Office of Basic Energy Sciences, of the U.S. Department of Energy under Contract No. DE-AC02-05CH11231.


\begin{thebibliography}{10}
\providecommand{\url}[1]{#1}
\csname url@samestyle\endcsname
\providecommand{\newblock}{\relax}
\providecommand{\bibinfo}[2]{#2}
\providecommand{\BIBentrySTDinterwordspacing}{\spaceskip=0pt\relax}
\providecommand{\BIBentryALTinterwordstretchfactor}{4}
\providecommand{\BIBentryALTinterwordspacing}{\spaceskip=\fontdimen2\font plus
\BIBentryALTinterwordstretchfactor\fontdimen3\font minus
  \fontdimen4\font\relax}
\providecommand{\BIBforeignlanguage}[2]{{%
\expandafter\ifx\csname l@#1\endcsname\relax
\typeout{** WARNING: IEEEtran.bst: No hyphenation pattern has been}%
\typeout{** loaded for the language `#1'. Using the pattern for}%
\typeout{** the default language instead.}%
\else
\language=\csname l@#1\endcsname
\fi
#2}}
\providecommand{\BIBdecl}{\relax}
\BIBdecl

\bibitem{Sun:2006bo}
X.~Sun, B.~Yu, G.~Ng, T.~D. Nguyen, and M.~Meyyappan, ``{III-VI compound
  semiconductor indium selenide (In$_2$Se$_3$) nanowires: Synthesis and
  characterization},'' \emph{Applied Physics Letters}, vol.~89, no.~23, p.
  233121, 2006.

\bibitem{Rasmussen:2013hz}
A.~M. Rasmussen, S.~T. Teklemichael, E.~Mafi, Y.~Gu, and M.~D. McCluskey,
  ``{Pressure-induced phase transformation of In$_2$Se$_3$},'' \emph{Applied
  Physics Letters}, vol. 102, no.~6, p. 062105, 2013.

\bibitem{Zhu:2010jj}
H.~Zhu, J.~Yin, Y.~Xia, and Z.~Liu, ``{Ga$_2$Te$_3$ phase change material for
  low-power phase change memory application},'' \emph{Applied Physics Letters},
  vol.~97, no.~8, p. 083504, 2010.

\bibitem{Kurosaki:2008cj}
K.~Kurosaki, S.~Yamanaka, and M.~Ishimaru, ``{Unexpectedly low thermal
  conductivity in natural nanostructured bulk Ga$_2$Te$_3$},'' \emph{Applied
  Physics Letters}, vol.~93, no.~1, 2008.

\bibitem{Luo:2004vw}
M.~Luo and M.~Wuttig, ``{The Dependence of Crystal Structure of Te Based Phase
  Change Materials on the Number of Valence Electrons},'' \emph{Advanced
  Materials}, vol.~16, no.~5, pp. 439--443, 2004.

\bibitem{Lencer:2011bt}
D.~Lencer, M.~Salinga, and M.~Wuttig, ``{Design Rules for Phase-Change
  Materials in Data Storage Applications},'' \emph{Advanced Materials},
  vol.~23, no.~18, pp. 2030--2058, 2011.

\bibitem{Warren:1974ux}
W.~W. Warren, ``{Solid immiscibility and liquid structure in the
  Ga$_2$(Se$_x$Te$_{1-x}$)$_3$ alloy system},'' \emph{Journal of Physics and
  Chemistry of Solids}, vol.~35, no.~9, pp. 1153--1157, 1974.

\bibitem{Krbal:2009in}
M.~Krbal, A.~Kolobov, J.~Haines, P.~Fons, C.~Levelut, R.~Le~Parc, M.~Hanfland,
  J.~Tominaga, A.~Pradel, and M.~Ribes, ``{Initial Structure Memory of
  Pressure-Induced Changes in the Phase-Change Memory Alloy
  Ge$_2$Sb$_2$Te$_5$},'' \emph{Physical Review Letters}, vol. 103, no.~11, p.
  115502, 2009.

\bibitem{AbdulJabbar:2012bma}
N.~Abdul-Jabbar, E.~D. Bourret-Courchesne, and B.~D. Wirth, ``{Single crystal
  growth of Ga$_2$(Se$_x$Te$_{1-x}$)$_3$ semiconductors and defect studies via
  positron annihilation spectroscopy},'' \emph{Journal of Crystal Growth}, vol.
  352, no.~1, pp. 31--34, 2012.

\bibitem{Kunz:2005ga}
M.~Kunz, A.~A. MacDowell, W.~A. Caldwell, D.~Cambie, R.~S. Celestre, E.~E.
  Domning, R.~M. Duarte, A.~E. Gleason, J.~M. Glossinger, N.~Kelez, D.~W.
  Plate, T.~Yu, J.~M. Zaug, H.~A. Padmore, R.~Jeanloz, P.~A. Alivisatos, and
  S.~M. Clark, ``{A beamline for high-pressure studies at the Advanced Light
  Source with a superconducting bending magnet as the source},'' \emph{Journal
  of Synchrotron Radiation}, vol.~12, no.~5, pp. 650--658, 2005.

\bibitem{Heinz:1984vw}
D.~L. Heinz and R.~Jeanloz, ``{The equation of state of the gold calibration
  standard},'' \emph{Journal of Applied Physics}, vol.~55, no.~4, pp. 885--893,
  1984.

\bibitem{Cellref}
\BIBentryALTinterwordspacing
J.~Laugier and B.~Bochu. (2002) computer code celref. version 3. cell parameter
  refinement program from powder diffraction diagram (laboratoire des materiaux
  et du genie physique, ecole nationale superieure de physique de grenoble
  [inpg]. [Online]. Available: \url{http://www.inpg.fr/LMGP}
\BIBentrySTDinterwordspacing

\bibitem{Kim:2011kb}
C.-E. Kim, K.~Kurosaki, M.~Ishimaru, H.~Muta, and S.~Yamanaka, ``{Effect of
  Vacancy Distribution on the Thermal Conductivity of Ga$_2$Te$_3$ and
  Ga$_2$Se$_3$},'' \emph{Journal of Electronic Materials}, vol.~40, no.~5, pp.
  999--1004, 2011.

\bibitem{AbdulJabbar:2014fl}
N.~M. Abdul-Jabbar, P.~Ercius, R.~Gronsky, E.~D. Bourret-Courchesne, and B.~D.
  Wirth, ``{Probing the local environment of two-dimensional ordered vacancy
  structures in Ga$_2$SeTe$_2$ via aberration-corrected electron microscopy},''
  \emph{Applied Physics Letters}, vol. 104, no.~5, p. 051904, 2014.

\bibitem{Guymont:1992vj}
M.~Guymont, A.~Tomas, and M.~Guittard, ``{The structure of Ga$_2$Te$_3$ an
  x-ray and high-resolution electron microscopy study},'' \emph{Philosophical
  Magazine A}, vol.~66, no.~1, pp. 133--139, 1992.

\bibitem{Kim:2009dz}
C.-E. Kim, K.~Kurosaki, M.~Ishimaru, D.-Y. Jung, H.~Muta, and S.~Yamanaka,
  ``{Effect of periodicity of the two-dimensional vacancy planes on the thermal
  conductivity of bulk Ga$_2$Te$_3$},'' \emph{physica status solidi (RRL) -
  Rapid Research Letters}, vol.~3, no. 7-8, pp. 221--223, 2009.

\bibitem{Huang:2013ke}
G.-Y. Huang, N.~M. Abdul-Jabbar, and B.~D. Wirth, ``{First-principles study of
  the structure and band structure of Ga$_2$Se$_3$},'' \emph{Journal of
  Physics-Condensed Matter}, vol.~25, no.~22, p. 225503, 2013.

\bibitem{Huang:2013jv}
G.-Y. Huang and B.~D. Wirth, ``{Energetics and kinetics of native point defects
  in Ga$_{2}$Se$_{3}$ from first principles},'' \emph{Physical Review B},
  vol.~88, no.~8, p. 085203, 2013.

\bibitem{Huang:2014vt}
G.-Y. Huang, N.~M. Abdul-Jabbar, and B.~D. Wirth, ``{Theoretical study of
  Ga$_2$Se$_3$, Ga$_2$Te$_3$ and Ga$_2$(Se$_{1-x}$Te$_x$)$_3$: Band-gap
  engineering},'' \emph{Acta Materialia}, vol.~7, pp. 349--369, 2014.

\bibitem{Perdew:1981dv}
J.~P. Perdew, ``{Self-interaction correction to density-functional
  approximations for many-electron systems},'' \emph{Physical Review B},
  vol.~23, no.~10, pp. 5048--5079, 1981.

\bibitem{Kresse:1993bz}
G.~Kresse and J.~Hafner, ``{Ab initio molecular dynamics for liquid metals},''
  \emph{Physical Review B}, vol.~47, no.~1, pp. 558--561, 1993.

\bibitem{Kresse:1996kl}
G.~Kresse, ``{Efficient iterative schemes for ab initio total-energy
  calculations using a plane-wave basis set},'' \emph{Physical Review B},
  vol.~54, no.~16, pp. 11\,169--11\,186, 1996.

\bibitem{Woolley:1959tl}
J.~C. Woolley and B.~R. Pamplin, ``{The ordered crystal structure of
  In$_2$Te$_3$},'' \emph{Journal of the Less Common Metals}, vol.~1, pp.
  362--376, 1959.

\bibitem{Lubbers:1982wd}
D.~L\"{u}bbers and V.~Leute, ``{The crystal structure of
  $\beta$--Ga$_2$Se$_3$},'' \emph{Journal of Solid State Chemistry}, vol.~43,
  no.~3, pp. 339--345, 1982.

\end{thebibliography}
\end{document}